\documentclass{svjour3}

\usepackage{amsmath, amssymb, epsfig, subfigure, a4wide}

\newcommand{\rd}{\mathrm{d}} 
\newcommand{\re}{\mathrm{e}} 
\DeclareMathOperator\sP{P}   
\DeclareMathOperator\sE{E}   

\journalname{Journal of Mathematical Biology}

\begin{document}


\title{How T-cells use large deviations to recognize foreign antigens}
\author{Natali Zint \and Ellen Baake \and Frank den Hollander}
\institute{N. Zint \and E. Baake \at
              Technische Fakult\"at, Universit\"at Bielefeld, Postfach 100131, 33501 Bielefeld, Germany \\
              \email{\{ebaake,nzint\}@techfak.uni-bielefeld.de}  
           \and
           F. den Hollander \at
              Mathematisch Instituut, Universiteit Leiden, Postbus 9512, 2300 RA Leiden, The Netherlands \&
	      EURANDOM, Postbus 513, 5600 MB Eindhoven, The Netherlands \\
	      \email{denholla@math.leidenuniv.nl}
}
\date{Received: date / Accepted: date}

\maketitle

\begin{abstract}
A stochastic model for the activation of T-cells is analysed. T-cells are part of 
the immune system and recognize foreign antigens against a background of the body's 
own molecules. The model under consideration is a slight generalization of a model 
introduced by Van den Berg, Rand and Burroughs in 2001 \cite{vdb01}, and is capable of explaining how 
this recognition works on the basis of rare stochastic events. With the help of a 
refined large deviation theorem and numerical evaluation it is shown that, for a
wide range of parameters, T-cells can distinguish reliably between foreign antigens 
and self-antigens.
\end{abstract}


\section{Introduction}

The mammalian immune system relies critically on so-called
T-cells, which recognize foreign antigens and trigger an immune response.
The word ``antigen'' is derived from {\em anti}body {\em gen}erating
(indicating that antigens are molecules that can elicit an immune reaction).

Each individual is supplied with a large repertoire of different types of
T-cells (each defined by the special type of T-cell receptor exposed at 
its surface), and every type recognizes a certain repertoire of antigens.
This recognition, in turn, starts a signalling cascade, which induces an 
immune response that finally leads to the elimination of the antigen 
(Janeway et al.\ \cite{jane05}).

The T-cell repertoire of an organism must, on the one hand, recognize
{\em foreign} antigens in a reliable way; on the other hand, it must 
{\em not} respond to the body's {\em own} antigens, since this would 
elicit dangerous auto-immune reactions. How does this ``self-nonself 
distinction'' work?

This basic question of immunobiology has remained unanswered for a very 
long time. One fundamental difficulty lies in the fact that foreign antigens
and self-antigens are very similar in nature. Van den Berg, Rand and Burroughs
\cite{vdb01} (henceforth referred to as BRB) addressed this difficulty by
modelling the {\em probabilistic} nature of the interactions between T-cell
receptors and the antigens presented on the surface of so-called antigen-presenting
cells (APCs). A T-cell (with many copies of its given receptor) encounters these 
APCs (which carry a random mixture of antigens). By modelling these encounters
as random events, and taking the interaction kinetics between T-cell receptors
and antigens into account, BRB have shown that the T-cell repertoire can distinguish
reliably between APCs that carry foreign antigens and those that do not.

In order to validate the results of BRB, we reconsider their model and refine 
its analysis. In mathematical terms, the model boils down to computing the  
distribution of a large sum of independent but not identically distributed 
random variables. Since a T-cell response is a rare event (for a randomly 
chosen encounter), the tail of the distribution is relevant --
a situation that requires the use of {\em large deviation theory}.
Specifically, we need  the so-called {\em exact asymptotics}, 
as provided by the Bahadur-Rao theorem and a generalization
proved by Chaganty and Sethuraman \cite{chag93}.
With the help of this theorem, we find substantially 
elevated tail probabilities for the case where a foreign antigen is present in 
a fairly high copy number, relative to the self-background. Abundance of the 
foreign antigen is biologically realistic, since pathogens multiply within the 
body and swamp it with their antigens before an immune response is started. 
Furthermore, the requirement of a high copy number can be relaxed in a refined 
version of the model that includes a biological mechanism known as negative 
selection (see below).

The BRB paper \cite{vdb01} appeared in a biological context (with 
lots of immunological detail not easily accessible to the non-specialist) 
and therefore 
put little emphasis on mathematical detail. The aim of the present article is 
threefold. Firstly, we will make this fascinating piece of theoretical biology 
available to a more mathematical readership. To this end, we will streamline the 
modelling by formulating a set of explicit assumptions. Secondly, we will put 
the analysis on a solid mathematical basis by stating and applying the necessary 
large deviation result. This result holds under rather general conditions and 
therefore opens up interesting perspectives for further research. Thirdly, we 
will put forward and analyse numerically an extension of the BRB-model obtained 
by replacing the constant copy numbers of antigens on APCs by random variables, 
which is biologically more realistic.

The remainder of this article (which builds on the thesis of Zint \cite{zint05}, 
where many more details may be found) is organized as follows. Section~\ref{s:biology} 
explains the immunological problem in a nutshell. 
Section~\ref{s:model} presents the mathematical model in its generalized
form, with the emphasis on making individual modelling steps and assumptions 
transparent. Section \ref{s:ld} is devoted to large deviations,
and provides the main theorem required for our analysis. On this basis,
approximations are derived in
Section \ref{s:activation_curves}, and applied to the biological model 
to demonstrate its recognition ability.
Section \ref{s:discussion}, finally, summarizes and discusses the results, the possible 
extensions, as well as the limitations of the model.


\section{T-cells and antigen recognition in a nutshell}\label{s:biology}

The object of immunobiology is the body's own defence against pathogens
like bacteria, viruses or fungi. One distinguishes between unspecific and
specific defence mechanisms. The latter form the so-called immune system, 
which specifically reacts to intruders. In this reaction, the T-cells play
an important role, which we will now briefly describe; for more details, see
the textbook by Janeway et al.\ \cite{jane05}.

\bigskip

{\em T-cells.}
T-cells are produced in the bone marrow and subsequently  migrate to the thymus,
where they mature (see below). On leaving the thymus, each T-cell is characterized 
by a specific type of T-cell receptor (TCR), which is displayed in many {\em identical} 
copies on the surface of the particular T-cell (see Fig.\ \ref{f:illustration}).
These TCRs play an important part in the recognition of intruders (see below). It 
is important to note that all TCRs on one T-cell are of the same type. However, a 
large number (roughly $10^7$; see Arstila et al.\ \cite{arst99}) of different receptors, 
and hence different T-cell types, are present in an individual.

\begin{figure}[h]
\centerline{\epsfig{file=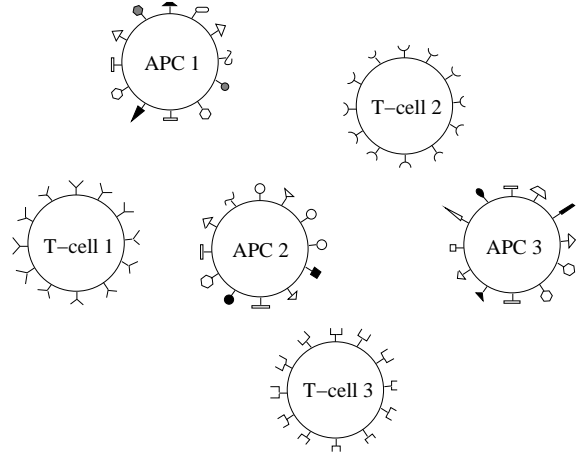,width=75mm}}
\caption{\label{f:illustration} A sample of different T-cells and APCs}
\end{figure}

\bigskip

{\em Antigen-presenting cells.}
The partners of the T-cells are the antigen-presenting cells (APCs), each producing 
so-called MHC molecules of different types (the set of these different types is the 
same for all the APCs of a given individual). An APC absorbs antigens from its vicinity 
and breaks them down. In the cell the emerging fragments, so-called peptides (short 
sequences of amino acids), are bound to the MHC molecules. The resulting complexes, 
composed of an MHC molecule and a peptide (abbreviated by pMHC), are displayed on 
the surface of the cell (the MHC molecules serve as ``carriers'' or ``anchors''
to the cell surface). Since most of the peptides in the vicinity of an APC are the 
body's own peptides, every APC displays a large variety of different types of 
self-peptides and, possibly, one (or a small number of) foreign types. The various 
types of peptides occur in various copy numbers, as will be detailed below. For the 
moment, we merely note that foreign peptides are often present at elevated copy 
numbers. As noted above, this is because pathogens multiply within the body and flood 
it with their antigens, before an immune response is initiated.

\bigskip

\begin{figure}[h]
\centerline{\epsfig{file=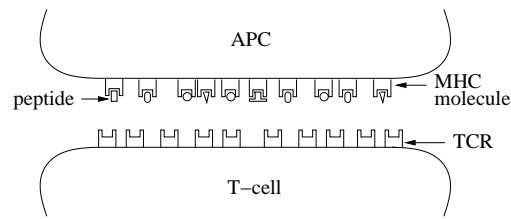,width=75mm}}
\caption{\label{f:synapse} An immunological synapse}
\end{figure}

{\em Interactions between T-cells and APCs.}\label{ss:reactions} 
The presentation of peptides on the surface of the APCs is of great importance
for the immune system, because T-cells will only be activated when they recognize
foreign peptides on the surface of an APC. The contact between a T-cell and an
APC is established by a temporary bond between the cells, through which a so-called
immunological synapse (see Fig.\ \ref{f:synapse}) is formed, in which the TCRs and the
pMHCs interact with each other. If a T-cell recognizes a foreign peptide through
its receptors, then it is activated to reproduce, and the resulting clones of T-cells 
will initiate an immune reaction against the intruder.

\vskip 0.3truecm

{\em Maturation.}
During the maturation of a T-cell, several processes take place in the thymus. Initially, 
the T-cell starts to display the TCRs on its surface. After this, two selective processes
take place. During positive selection, those T-cells that hardly interact with the MHC 
molecules of the individual are removed. Furthermore, negative selection causes the removal
of those T-cells that react too strongly to self-peptides. Thus, both useless and dangerous 
T-cells are removed.

\bigskip

{\em Problem.}
Self-peptides and foreign peptides cannot differ from each other by nature. After all, even 
tissues of a different individual of the same species are recognized as foreign (this is the 
basic problem of transplantation). However, the activation of a T-cell occurs only when it 
recognizes a foreign peptide. Therefore the question comes up how the T-cells can distinguish 
between self and non-self. At first sight, the task seems hopeless, since there are vastly more
different peptides (roughly $10^{13}$; see Mason \cite{maso98}) than TCRs (roughly $10^7$, as
noted earlier), which makes specific recognition (where one TCR recognizes exactly one pMHC) 
impossible; this is known as the Mason paradox. Fortunately, there is an answer to this question.


\section{The model} 
\label{s:model}

We present here a slightly generalized version of the model originally proposed in 
BRB \cite{vdb01}. To this end, we recapitulate the modelling ideas and idealizations, 
and summarize them as assumptions (A1)--(A7) below.

Consider the immunological synapse between a T-cell and an APC (see Fig.\ \ref{f:synapse}).
The activation of the T-cell is conceived as follows. If a contact between a TCR and 
a pMHC lasts longer than a certain time period, $t^*$, then the T-cell will receive a 
stimulus. The T-cell adds up the stimulation rates of all its receptors. If this sum 
exceeds a threshold, $g_{\rm act}$, then the T-cell will be activated. This model 
relies on several hypotheses, which  are known as kinetic 
proofreading (\cite{coom02},\cite{mcke95}), serial triggering (\cite{vali95},\cite{vali97}) 
and counting of stimulated TCRs (\cite{roth96},\cite{viol96}).

\subsection{Kinetics of stimulation}

Let ${\rm R}_i$ be an unbound TCR of type $i$, ${\rm M}_j$ an unbound pMHC of type
$j$, and ${\rm C}_{ij}$ a complex composed of ${\rm R}_i$ and ${\rm M}_j$, where
$i,j \in \mathbb{N}$. For every such pair, the binding and unbinding may, in chemical 
shorthand notation, be symbolized as
\begin{equation}\label{G:Reaktion}
{\rm R}_i+{\rm M}_j\overset{\lambda_{ij}}{\underset{\rho_{ij}}{\rightleftharpoons}} 
{\rm C}_{ij}\; ,
\end{equation}
where $\lambda_{ij}$ and $\rho_{ij}$ are the association and dissociation rates, 
respectively. An encounter (to be used synonymously with an immunological synapse) 
between a T-cell and an APC with its mixture of pMHCs is therefore characterized by 
the type $i$ of the T-cell, the types $j$ of pMHCs at hand and the associated surface 
densities $z_j$, as well as the rates $\lambda_{ij}$ and $\rho_{ij}$ (which will be
considered fixed for the purpose of this subsection). 
If the spatial structure in the
immunological synapse is ignored, then the corresponding kinetics in the synapse is
described by the deterministic law of mass action, 
i.e., for the given $i$, 
\begin{equation}\label{G:Dgl}
\begin{aligned}
\frac{\rd}{\rd t} c_{ij}(t) &= \lambda_{ij} r_i(t)m_j(t)-\rho_{ij} c_{ij}(t)\; ,
& c_{ij}(0)&=0\; , & \forall j\; ,\\
\frac{\rd}{\rd t} m_j(t) &= -\lambda_{ij} r_i(t)m_j(t)+\rho_{ij} c_{ij}(t)\; ,
& m_j(0)&= z_j\; , & \forall j \; ,\\
\frac{\rd}{\rd t} r_i(t) &= -\sum_j \lambda_{ij} r_i(t)m_j(t)+ \sum_j \rho_{ij} c_{ij}(t)\; ,
& r_i(0)&= r\; ,&
\end{aligned}
\end{equation}
where $r_i(t)$, $m_{j}(t)$ and $c_{ij}(t)$ are the surface densities of ${\rm R}_i$, 
${\rm M}_j$ and ${\rm C}_{ij}$, respectively, at time $t$ (and we note that only
finitely many $z_j$'s are non-zero). It is easily verified that the solution of 
(\ref{G:Dgl}), for the given $i$, 
satisfies the conservation laws
\begin{equation}\label{G:Con_law}
r_i(t) + \sum_j c_{ij}(t)=r \qquad \text{and} \qquad m_j(t)+ c_{ij}(t) = z_j 
\quad \forall j\;  .
\end{equation}

Note that this deterministic approach, with its surface densities varying over
the reals rather than a finite set, is justified when the numbers of all the 
involved molecules are large enough (see e.g.\ Ethier and Kurtz \cite[Chapter 11,
Theorem 2.1]{ethi86}). This  standard approach to reaction kinetics
is also used for the binding kinetics in the immunological synapse 
(see e.g.\ BRB \cite[Eq.\ (A.5)]{vdb02}, which describes the equilibrium of this 
model).
It should be noted, however, that some of the antigen types may be fairly
rare in the situation at hand, so that the deterministic approach may be 
somewhat crude. But it will become clear later on that we actually do not 
depend on details of the binding kinetics.

\noindent
[Remark: In the probabilistic approach in the next subsection, we will use the symbol 
$z_j$ for the copy number of type $j$ pMHC rather than its surface density,
since they differ by an (irrelevant) normalization factor only.]

\bigskip

{\em Equilibrium.}
The bond between a TCR and a pMHC consists of two parts: contacts between the TCR 
and the MHC molecule, and contacts between the TCR and the peptide. According to 
Wu et al.\ \cite{wu}, the former specify mainly the association rate and the 
latter mainly the dissociation rate. We consider mature T-cells, which implies 
that they have been positively selected, i.e., each T-cell binds one type of the 
MHC molecules of the individual very well. Thus, we idealize $\lambda_{ij}$ as 
very large ($\gg 1/(r\,t^*)$) for pMHCs containing this type of MHC molecule, and 
zero otherwise. For every $i$, we therefore restrict $j$ in (\ref{G:Dgl}) to the 
set
\begin{equation} \label{e:restriction}
\mathcal{P}_i =\{j \colon\, \lambda_{ij} \gg 1/(r\,t^*) \}\; ,
\end{equation}
since only these pMHCs contribute significantly to the stimulation rate. As a consequence,
we may assume that the reaction is in equilibrium, because this is reached on a time scale 
that is short relative to the time scale of activation. Therefore, in view of (\ref{G:Dgl}),
for every $i$ we have
\begin{equation} \label{G:equi}
\hat{c}_{ij} = \frac{\lambda_{ij}}{\rho_{ij}}\,\hat{r}_i\hat{m}_j
\qquad \forall\,j \in \mathcal{P}_i\; ,
\end{equation}
where $\hat{r}_i$, $\hat{m}_j$ and $\hat{c}_{ij}$ denote the equilibrium quantities.
Combining (\ref{G:Con_law}) for these quantities with (\ref{G:equi}), we get an equation 
for $\hat{c}_{ij}$ that can be solved to give the implicit system of equations
\begin{equation} \nonumber
\hat{c}_{ij} = z_j \,\frac{r- \left( \sum_{k \in \mathcal{P}_i} \hat{c}_{ik} \right)}
{r - \left( \sum_{k \in \mathcal{P}_i} \hat{c}_{ik} \right) + \rho_{ij}/\lambda_{ij}}
\end{equation}
(cf.\ BRB \cite[Appendix A.1]{vdb02}). Assuming further that the concentration of TCRs 
is not limiting (i.e., $r > \sum_{j \in \mathcal{P}_i} z_j$ -- this is the so-called 
serial triggering regime) and that the relevant dissociation rates are very small (i.e.,
$\rho_{ij}/\lambda_{ij} \ll r$ for all $j \in \mathcal{P}_i$), we are led to the idealization

\begin{itemize}
\item[(A1)]\hfill
  $\hat{c}_{ij} = z_j\; .$\hfill \hphantom{x}
\end{itemize}

\noindent
As will become clear later on, this assumption may actually be relaxed (see 
Section \ref{s:discussion}); we make it here for ease of exposition.

\bigskip

{\em Stimulation rate.}
A given ${\rm C}_{ij}$ dissociates at rate $\rho_{ij}$. Therefore, $T_{ij}$, the 
duration of a contact between ${\rm R}_i$ and ${\rm M}_j$, is exponentially 
distributed with mean $\tau_{ij}=1/\rho_{ij}$. Hence, the probability of $T_{ij}$ 
exceeding $t^*$ is $\re^{-t^*/\tau_{ij}}$, which we refer to as the {\em stimulation probability}
(of a ${\rm C}_{ij}$). Note that, by the above together with 
(\ref{e:restriction}) and (A1), an encounter between a T-cell and an APC is 
now characterized by $i$, $\mathcal{P}_i$ as well as the collections of $z_j$ and
$\tau_{ij}$ for all $j \in \mathcal{P}_i$ with $z_j>0$.

The T-cell receives a stimulus every time a complex dissociates that has existed 
for at least time $t^*$. Therefore the average stimulation rate of type $ij$ 
complexes is given by

\begin{itemize}
\item[(A2)] \hfill
  $\rho_{ij} \sP(T_{ij}>t^*)=w(\tau_{ij}) \quad \mbox{with} \quad
  w(\tau)=\frac{1}{\tau}\re^{\frac{-t^*}{\tau}}\; .$\hfill \hphantom{x}
\end{itemize}

\begin{figure}[h]
\centerline{\epsfig{file=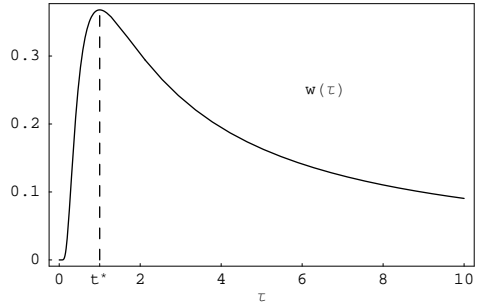,width=65mm}}
\caption{\label{f:stimulation_rate} Average stimulation rate of an individual 
complex type as a function of the average waiting time}
\end{figure}

\noindent
In Fig.\ $\ref{f:stimulation_rate}$, $w(\tau)$ is plotted as a function of $\tau$. 
This curve can be interpreted as follows.

\begin{enumerate}
\item If $\tau \ll t^*$, then the complex will typically dissociate before stimulation.
\item If $\tau \gg t^*$, then the TCR and the pMHC will typically be associated for 
a long time. Therefore the T-cell will get a stimulus through practically every 
binding event, but the pMHC keeps the receptor occupied for a long time, so only 
few stimuli are expected per time unit.
\end{enumerate}

By (A1) and (A2), we finally get the total stimulation rate for a conjunction of 
a T-cell of type $i$ and a particular APC:
\begin{itemize}
\item[(A3)]\hfill
$g_i=\sum_{j \in \mathcal{P}_i} z_j w(\tau_{ij})\; .$\hfill \hphantom{x}
\end{itemize}

\subsection{A probabilistic approach}

So far, $\mathcal{P}_i$, $z_j$ and $\tau_{ij}$ are considered to be given quantities
for all $i,j\in \mathbb{N}$. Indeed, $z_j$ and $\tau_{ij}$ may be determined 
experimentally for a given $i$, $j$ pair (cf.\ \cite{hun92},\cite{mat94}). However, 
owing to the diversity of complexes $C_{ij}$ and mixtures of peptides presented 
on the APCs, it is not possible to specify all these quantities individually. 
Therefore, in order to derive the overall probability of T-cell activation, a 
probabilistic approach is required.

\bigskip

{\em Presentation of antigens.}
The genes of an organism can be classified as constitutive ones and inducible ones.
The former encode proteins that are always present in every cell (e.g.\ 
proteins of the basic
metabolism). In contrast, the latter encode proteins that only exist in some cells
(like for example muscle proteins) and/or occur only temporarily, i.e., they are variable.
Accordingly, the types of self-peptides on each APC may be partitioned into constitutive 
and variable, i.e., $\mathcal{P}_i=\mathcal{C}_i \cup \mathcal{V}_i$, where 
$\mathcal{C}_i \cap \mathcal{V}_i = \emptyset$, and we suppose that:
\begin{itemize}
\item[(A4)]
There are constant numbers $n_c$ and $n_v$ of constitutive and variable types of 
peptides, respectively, on each APC.
\end{itemize}
The constitutive types ($\mathcal{C}_i$) are the same on each APC, whereas there is 
a different sample of variable types ($\mathcal{V}_i$) on each APC. As a generalization 
of BRB \cite{vdb01} and Zint \cite{zint05}, we allow the copy numbers of the individual 
types within each class to vary. Therefore we suppose that: 
\begin{itemize}
\item[(A5)]
The $z_j$ are realizations of random variables denoted by $Z_j$. These random 
variables are independent and identically distributed (i.i.d.) within each of the two classes,
and are referred to as $Z_j^{(c)}$ and $Z_j^{(v)}$.
\end{itemize}
[Remark: The i.i.d.\ assumption is made for simplicity; we do not model a particular biological
mechanism here. Realistic models would be based on MHC loading fluctuations 
(cf.\ BRB \cite[Appendix C]{vdb01}). They invariably induce dependencies, the
treatment of which is beyond the scope of the present paper.]

Let us now add in foreign peptides, and make the simplifying assumption that only one
type of foreign peptide is present on an APC, in $z_f$ copies. In the following, the 
index set for the pMHCs is therefore the union of $\mathbb{N}$ and $\{f\}$.

The total stimulation rate with respect to a conjunction of a T-cell of type $i$ and a
randomly chosen APC is given by
\begin{equation}\nonumber
G_i(z_f)=\left( \sum_{j \in \mathcal{C}_i} q Z_j^{(c)} w(\tau_{ij}) \right) +
\left( \sum_{j \in \mathcal{V}_i} q Z_j^{(v)} w(\tau_{ij}) \right) + z_f w(\tau_{if})\; .
\end{equation}
Here, the factor $q=(n_M-z_f)/n_M$ ensures that adding the foreign peptides does not 
change the expected number $n_M=n_c \sE(Z_1^{(c)}) +  n_v \sE(Z_1^{(v)})$ 
of MHC 
molecules on the surface of an APC, since 
also $q (n_c \sE(Z_1^{(c)}) +  n_v \sE(Z_1^{(v)})) +z_f =n_M$.

\bigskip

{\em Probability of activation.}
In line with the original model, we suppose that:
\begin{itemize}
\item[(A6)]
For all $i$ and $j$, the $\tau_{ij}$ are realizations of i.i.d.\ random variables
$\mathcal{T}_{ij}$ with mean $\bar{\tau}$ (see BRB \cite{vdb01} and Zint \cite{zint05} 
for explanations).
\end{itemize}
In particular, this assumption means that no distinction between foreign and self is 
built into the interaction between receptors and antigens. This reflects the fact 
that there is no a-priori difference between the peptides.

Note that (A6) also implies that the $\mathcal{P}_i$ need not be specified and 
the index $i$ may be suppressed, because we consider an arbitrary T-cell and the 
$G_i$ are i.i.d.\ random variables. Biologically, this means that
we choose a 
new T-cell type (as well as a new APC) for each encounter, 
and ignore further meetings of T-cells of the same type.
As a consequence,
the partitioning into constitutive and variable peptides has, at this stage, no 
effect except for the different abundances.

Altogether the total stimulation rate with respect to a conjunction of a randomly 
chosen T-cell and an APC is given by
\begin{equation} \label{G:ZV_Trigger_fremd}
G(z_f)= \left( \sum_{j=1}^{n_c} q Z_j^{(c)} W_j \right) + 
\left( \sum_{j=n_c+1}^{n_c+n_v} q Z_j^{(v)} W_j \right) + z_fW_{n_c+n_v+1} \; ,
\end{equation}
where $W_j=w(\mathcal{T}_j)$, and:
\begin{itemize}
\item[(A7)]
The probability of T-cell activation is $\sP \left(G(z_f) \geq g_{\rm act}\right) $.
\end{itemize}

{\em Specification of distributions and parameter values.}
For ease of exposition (and in line with the original model), $\mathcal{T}_{ij}$ 
is exponentially distributed and we have chosen $t^*=1$, $\bar{\tau}=0.04\,t^*$, 
$n_c=50$ and $n_v=1500$. As stated in BRB \cite{vdb01}, the number of MHC molecules 
ranges between $10^4$ and $10^6$. Therefore we take $n_M=10^5$. Furthermore, we 
use binomial distributions ${\rm Bin}_{m_c,p}$ and ${\rm Bin}_{m_v,p}$ for 
$Z_j^{(c)}$ and $Z_j^{(v)}$ , for all $j$, 
with parameters $m_c=1000$, $m_v=100$ and $p=0.5$,
so that the means $\sE(Z_1^{(c)})= 500$ and $\sE(Z_1^{(v)})=50$ 
correspond to 
the values $z_c/n_M= 0.005$ and $z_v/n_M=0.0005$ in the BRB-model. (Apart from 
the expectation, the distribution is an ad-hoc choice.)

It should be mentioned that moderate changes of the values of the parameters
(like for example $n_c$, $n_v$, $\sE(Z_1^{(c)})$ and $\sE(Z_1^{(v)})$) do not qualitatively
alter the results. Furthermore, these values have been chosen on the grounds of experimental
data (see BRB \cite{vdb01} and Zint \cite{zint05}). The exponential distribution is less-well founded. To show that
the qualitative behavior does not rely on this particular distribution, we will, as an alternative,
use the log-normal distribution (with parameters $\mu=-3.3$ and $\sigma=0.5$) in Section~\ref{s:actneg}.
This has a justification in terms of binding/unbinding kinetics (see BRB \cite{vdb01} and Zint \cite{zint05}).

\subsection{Distinction between foreign and self}

For the immune system to work, two conditions are essential: (a) if a foreign
antigen is present, then at least one T-cell will be activated; (b) there will 
be no activation when only self-antigens are present.

It is helpful to recast this into a hypothesis testing framework, in
the following way. The immune system performs a test of the null
hypothesis
\begin{equation}\label{eq:H_0}
    H_0: \; z_f = 0
\end{equation}
against the alternative hypothesis
\begin{equation}\label{eq:H_A}
    H_A: \; z_f >0\,.
\end{equation}
The test is performed via $N$ independent encounters between a T-cell and
an APC. $H_0$ is then rejected (and $H_A$
assumed) if at least one encounter leads to the event $\{G \geq g_{\rm act}\}$;
otherwise, $H_0$ is retained. The type I error is therefore
\begin{equation}\nonumber
  \alpha = \sP(H_A \; \text{assumed} \mid H_0 \; \text{true}) =
  1 - \left(1-\sP\left(G(0) \geq g_{\rm act}\right)\right)^N,
\end{equation}
and the type II error  is 
\begin{equation}\label{beta}
  \beta = \sP(H_0 \; \text{assumed} \mid  H_A \; \text{true}) =
   \left(1-\sP\left(G(z_f) \geq g_{\rm act}\right)\right)^N\,.
\end{equation}
Here, the underlying assumptions are (A1)--(A7), and
$G(z_f)$ is the sum of random variables introduced in Equation
\eqref{G:ZV_Trigger_fremd}, with constants $n_c$, $n_v$ and parameter $z_f$. In 
particular, $G(0)$ denotes the total stimulation rate in the absence of foreign 
peptides. The parameter $g_{\rm act}$  can be fine-tuned by the 
cell; for more on activation threshold tuning, see Van den Berg and Rand \cite{vdb04_2}.

Clearly, $\alpha$ is the probability of an autoimmune
response, whereas $\beta$ is the probability that a foreign antigen
goes unnoticed. By (b) and (a) above, both $\alpha$ and $\beta$ must
be small for the self-nonself distinction to work. 

Both $\sP(G(0) \geq g_{\rm act})$ and
$\sP(G(z_f) \geq g_{\rm act})$ are rare events (since at most a tiny fraction
of the T-cells reacts to a given APC).
Therefore, $\alpha$ is close to $0$ (close to $1$) and 
$\beta$ is close to $1$ (close to $0$) for $N$ small ($N$ very large).
We have no good knowledge of the value of $N$ (except that it is bounded
above by the number of T-cell types). But it is clear that a necessary
condition for distinction is that
$g_{\rm act}$ can be chosen in such a way that, for physiologically 
realistic values of $z_f$, 
\begin{itemize}
\item[(C1)]\hfill
$\sP (G(z_f) \geq g_{\rm act}) \gg \sP (G(0) \geq g_{\rm act}).
$\hfill \hphantom{x}
\end{itemize}
Consequently, there is a region of intermediate values of $N$ for which
both $\alpha$ and $\beta$ are small.
We thus need an analysis of the tiny probabilities
$\sP(G(0) \geq g_{\rm act})$ and $\sP(G(z_f) \geq g_{\rm act})$. This requires
large deviation theory and is
the subject of Section 4.


\section{Large deviations}
\label{s:ld}

For many families of random variables, {\em large deviation principles} 
(LDP's) are available to characterize their atypical behaviour.
Here, we will be concerned with sums 
of random variables, i.e., 
\[
S_n = \sum^n_{i=1} X_i,
\]
where $(X_i)_{i\geq 1}$ is a sequence of independent 
(but {\em not} necessarily identically distributed)
random variables (like those in Eq.~\eqref{G:ZV_Trigger_fremd}).
An LDP characterizes the probability of a large deviation of $S_n$ from its 
expectation; a {\em large deviation} is a deviation of order $n$ (in
contrast to a {\em normal deviation} of order $\sqrt{n}$, as 
covered by the central limit theorem). A basic result is Cram{\'e}r's 
theorem, which says the following. For a sequence $(X_i)_{i\geq 1}$ 
of i.i.d.\ real-valued random variables whose moment-generating function 
$\phi (\vartheta ) = \sE (\exp (\vartheta X_1))$ is finite
for all $\vartheta \in \mathbb{R}$, one has, for all $a> \sE(X_1)$,
\begin{equation}
\lim_{n\to\infty} \frac{1}{n} \log \sP (S_n \geq a n ) = - I(a),
\label{eq:cramer}
\end{equation}
where $I(a) = a \vartheta_a - \psi (\vartheta_a )$,  $\psi 
(\vartheta ) = \log \phi (\vartheta)$, 
and $\vartheta_a$ is the (unique) solution of $\psi'(\vartheta )=a$. 
That is, for large $n$, the probability that $S_n$ is larger than $a n$ 
decays exponentially with $n$, with decay rate 
$I(a)$. The value $\vartheta_a$ is known as the \emph{``tilting'' parameter}. 
It is used for an exponential reweighting (or ``tilting'') 
of the distribution 
of the $X_i$ (and hence of $S_n$) that inflates the right-hand tail of
the distribution in such a way that the rare event $\{S_n \geq an\}$ turns
into a typical one; this is a crucial step in the analysis. For
a review of large deviation theory, see  e.g.\ Den Hollander \cite{dh00};
Cram{\'e}r's theorem and its proof are found in Ch.~I.3.

Note, however, that the knowledge of the exponential decay rate $I(a)$ 
alone does 
not suffice to provide meaningful leading-order estimates  of the 
probabilities of the rare event itself. This is because \eqref{eq:cramer} is compatible 
with $\sP (S_n \geq an) = f(n)\exp\left(-nI(a)(1+o(1))\right)$ for 
any prefactor $f(n) = \mathcal{O} (n^\alpha )$, with arbitrary $\alpha$. 
More accurate information is obtained from so-called {\em exact asymptotics}. 
For the situation at hand, this is given by a refinement of Cram{\'e}r's 
theorem due to Bahadur and Rao (cf.\ \cite{DZ94}, Ch.~3.7). 
Namely, under the assumptions 
of Cram{\'e}r's theorem and the additional requirement that the distribution 
of $X_1$ be non-lattice (which is always fulfilled if $X_1$ has a density), 
one has
\begin{equation}
\label{eq:BR}
\sP (S_n \geq an) = \frac{1}{\sqrt{2\pi n} \sigma \vartheta_a} 
e^{-nI(a)} (1+ o (1)) \mbox{ as } n\to \infty
\end{equation}
for all $a$ that satisfy $\sE (X_1) < a < \sup_\vartheta \psi 
(\vartheta )$. 
Here, $I(a)$ and $\vartheta_a$ are as above, and $\sigma^2 = 
\psi''(\vartheta_a)$ 
is the variance of $\frac{1}{n}S_n$ after ``tilting'' with the 
exponential parameter 
$\vartheta_a$. (The condition $a<\sup_{\vartheta} \psi (\vartheta )$ ensures 
that only those events $\{S_n \geq an\}$ are considered that are
actually possible; the condition is void if $X_1$, and hence
$S_n/n$, take values in all of $\mathbb{R}_{\geq 0}$.)

What we need to tackle our stimulation rates 
\eqref{G:ZV_Trigger_fremd}
is the generalization of \eqref{eq:BR} to situations 
in which the $(X_i)_{i\geq 1}$ are {\em not} identically distributed.
Fortunately, a very general result is available, 
which does not even require independence. This is the result of 
Chaganty and Sethuraman 
\cite{chag93},  which plays a crucial role in our analysis,
and which we will now formulate.

Let $(S_n)_{n \in \mathbb{N}}$ be a sequence of $\mathbb{R}$-valued random variables, 
with moment generating functions $\phi_n(\vartheta) = \sE\left(\exp(\vartheta S_n)\right)$, 
$\vartheta\in\mathbb{R}$. Suppose that there exists a $\vartheta^*\in (0,\infty)$ 
such that
\begin{equation}
\sup_{n\in\mathbb{N}}\, \sup_{\vartheta \in B_{\vartheta^*}} 
\phi_n(\vartheta) < \infty\; , 
\end{equation}
where $B_{\vartheta^*}=\{\vartheta\in\mathbb{C}\colon\,|\vartheta|<\vartheta^*\}$. 
Define
\begin{equation}\label{psiphi}
\psi_n(\vartheta) = \frac{1}{n} \log \phi_n(\vartheta)\; ,
\end{equation}
and let $(a_n)_{n\in\mathbb{N}}$ be a bounded sequence in $\mathbb{R}$ such that 
for each $n$ the equation
\begin{equation}
a_n = \psi'_n(\vartheta)
\end{equation}
has a solution $\vartheta_n \in (0,\vartheta^{**})$ for some $\vartheta^{**} \in 
(0,\vartheta^*)$. This solution is unique by strict convexity of $\psi_n$. Define
\begin{equation}
\begin{aligned}
\sigma_n^2 &= \psi''_n(\vartheta_n)\; ,\\
I_n(a_n) &= a_n \vartheta_n - \psi_n(\vartheta_n)\; . 
\end{aligned}
\end{equation}

\begin{theorem}[Chaganty-Sethuraman \cite{chag93}]
\label{T:verall_B_R}
If\/ $\inf_{n\in\mathbb{N}}\sigma_n^2>0$, $\lim_{n \to \infty} \vartheta_n\sqrt{n}
=\infty$ and
\begin{equation}\label{e:condition3}
\lim_{n\to\infty} \sqrt{n}\,\sup_{\delta_1 \leq |t| \leq \delta_2\vartheta_n}
\left|\frac{\phi_n(\vartheta_n+it)}{\phi_n(\vartheta_n)}\right| = 0 \quad
\forall\, 0<\delta_1<\delta_2<\infty\; ,
\end{equation}
then
\begin{equation}
\sP \left( S_n \geq n a_n \right) = 
\frac{\re^{-n I_{n}(a_n)}}{\vartheta_n \sigma_n \sqrt{2\pi n}}\,\left(1+o(1)\right)
\qquad \mbox{ as } n \to\infty\; .
\end{equation}
$\hspace*{\fill}\square$
\end{theorem}

In analogy with the previous discussion,
$\vartheta_n$ is the ``tilting parameter'' for the distribution of $\frac{1}{n}S_n$, 
$\sigma_n^2$ is the variance of the ``tilted'' $\frac{1}{n}S_n$, and $I_{n}(a_n)$ is
the large deviation rate function. Let us further remark that, in principle, 
even finer 
error estimates (of Berry-Ess\'een type, cf.\ \cite{fell71}, Ch.\ XVI)
can be obtained beyond the asymptotics in 
Theorem \ref{T:verall_B_R}, but this becomes technically more involved.


\section{Activation curves}
\label{s:activation_curves}

In order to investigate whether condition (C1) can be fulfilled for physiologically
realistic values of $z_f$, we consider the so-called activation curves, i.e., 
$1-F_{z_f}(g_{\rm act}$) with $F_{z_f}$ the distribution function of $G(z_f)$.

\subsection{Simulation and approximation}

We begin by deriving an approximation for the activation probability in condition (C1)
based on Theorem~\ref{T:verall_B_R}. Consider a sequence of models defined by increasing
numbers of constitutive and variable peptide types. Let
\begin{equation}\nonumber
n=
\begin{cases}
n_c + n_v, & \text{if $z_f=0$,}\\
n_c + n_v +1, & \text{otherwise,}
\end{cases}
\end{equation}
and consider the limit $n \to \infty$ with $\lim_{n \to \infty} n_c/n_v = C_1 \in 
(0,\infty)$; 
note that the number of foreign peptides remains 1 throughout.
Let $S_n$ in Theorem \ref{T:verall_B_R} be
\[
G_n(z_f)= \left( \sum_{j=1}^{n_c} q_n Z_j^{(c)} W_j \right) +
\left( \sum_{j=n_c+1}^{n_c+n_v} q_n Z_j^{(v)} W_j \right) + z_fW_{n_c+n_v+1}
\]
where
\[
q_n = \frac{n_c\,m_c\,p + n_v\,m_v\,p-z_f}{n_c\,m_c\,p + n_v\,m_v\,p}\; ,
\]
and let $M_c$, $M_v$ and $M$ be the moment generating functions of $Z_j^{(c)} W_j$, 
$Z_j^{(v)} W_j$ and $W_j$, respectively, i.e., for $\gamma \in \{ c,v\}$,
\begin{equation}\label{e:m1}
M_{\gamma}(\vartheta)=\frac{1}{\overline{\tau}} 
\sum_{k=0}^{m_{\gamma}} \left( \int_0^{\infty}
\exp\left( k \vartheta\, \frac{\exp(-t^*/\tau)}{\tau}-\frac{\tau}{\overline{\tau}}\right) 
\rd\tau\right) {\rm Bin}_{m_{\gamma},p}(k)\; ,
\end{equation}
and
\begin{equation}\label{e:m3}
M(\vartheta)=\frac{1}{\overline{\tau}} \int_0^{\infty}
\exp\left(\vartheta\, \frac{\exp(-t^*/\tau)}{\tau}-\frac{\tau}{\overline{\tau}}\right) \rd\tau\; .
\end{equation}
Choose $a_n\equiv a$ and $g_{\rm act}(n) = an$. Let $\vartheta_{n}$ be the unique 
solution of
\begin{equation}
\begin{aligned}\label{e:theta}
a &=& &\frac{n_c}{n} q_n \left[\frac{\rd}{\rd \vartheta} \log M_c(\vartheta)\right]
\Big\vert_{\vartheta=q_n\vartheta_n}\\
& &+  \, &\frac{n_v}{n} q_n \left[\frac{\rd}{\rd \vartheta} \log M_v(\vartheta)\right]
\Big\vert_{\vartheta=q_n\vartheta_n}
+ \, &\frac{1}{n} z_f \left[\frac{\rd}{\rd \vartheta} \log M(\vartheta)\right]
\Big\vert_{\vartheta=z_f\vartheta_n}\; .
\end{aligned}
\end{equation}
We further define
\begin{equation}
\begin{aligned}\label{e:sigma}
\sigma_{n}^2 &=&
& \frac{n_c}{n} q_n^2 \left[\frac{\rd^2}{\rd \vartheta^2} \log M_c(\vartheta)\right]
\Big\vert_{\vartheta=q_n\vartheta_n}\\
& & + \, &\frac{n_v}{n} q_n^2 \left[\frac{\rd^2}{\rd \vartheta^2} \log M_v(\vartheta)\right]
\Big\vert_{\vartheta=q_n\vartheta_n}
+\frac{1}{n} z_f^2 \left[\frac{\rd^2}{\rd \vartheta^2} \log M(\vartheta)\right]
\Big\vert_{\vartheta=z_f\vartheta_n}
\end{aligned}
\end{equation}
and
\begin{equation}\label{e:I}
I_{n}(a) = a \vartheta_n -\frac{n_c}{n} \log M_c(q_n\vartheta_n)-
\frac{n_v}{n} \log M_v(q_n\vartheta_n) - \frac{1}{n} \log M(z_f\vartheta_n) \; .
\end{equation}

Since we have only finitely many different types of random variables, all independent, 
it is straightforward to check 

\begin{lemma} \label{condokay}
The conditions for Theorem $\ref{T:verall_B_R}$ are satisfied.
\end{lemma}

\begin{proof}
The moment generating function $\phi_n(\vartheta)$ is given by
\begin{equation}\nonumber
\phi_n(\vartheta)=
\begin{cases}
\left(M_c(\vartheta)\right)^{n_c} \left(M_v(\vartheta)\right)^{n_v}, 
& \text{if $z_f=0$,}\\
\left(M_c(q_n \vartheta)\right)^{n_c} \left(M_v(q_n \vartheta)\right)^{n_v} 
M(z_f \vartheta),
& \text{otherwise.}
\end{cases}
\end{equation}
Let $z_f$ be fixed.
If $a$ is chosen such that $g_{\rm act}(n)>\sE(G_n(0))$ and $g_{\rm act}(n)>\sE(G_n(z_f))$ 
for all $n$ (in which case the strict inequalities are in fact uniform in $n$), then 
$\lim_{n\to\infty} \vartheta_n = C_2 \in (0,\infty)$. 
Consequently, $\lim_{n \to \infty} 
\vartheta_n \sqrt{n} = \infty$, and also $\inf_{n\in\mathbb{N}}\sigma_n^2>0$. It thus 
remains to verify Condition (\ref{e:condition3}). Let $F_c(x)$, $F_v(x)$ and $F(x)$ be 
the distribution functions of $Z_j^{(c)} W_j$, $Z_j^{(v)} W_j$ and $W_j$, respectively. 
Since
\[
\nu_{\gamma}^{(n)}(t)
=\frac{M_{\gamma}(q_n (\vartheta_n+it))}{M_{\gamma}(q_n \vartheta_n)}
=\int_{\mathbb{R}} \exp(i q_n t x)\, 
\frac{\exp(q_n \vartheta_n x)}{M_{\gamma}(q_n \vartheta_n)}\,  
dF_{\gamma}(x), \qquad \gamma \in \{ c,v\} \; ,
\]
and
\[
\nu^{(n)}(t)=\frac{M(z_f (\vartheta_n+it))}{M(z_f \vartheta_n)}
=\int_{\mathbb{R}} \exp(i z_f t x)\, 
\frac{\exp(z_f \vartheta_n x)}{M(z_f \vartheta_n)}\,  
dF(x)
\]
are characteristic functions of random variables that are not constant nor are 
lattice valued, and $\vartheta_n$ and $q_n$ converge as $n\to\infty$, there
exists an $\varepsilon>0$ and an $n_0<\infty$ such that, for all $t \neq 0$ 
and $n \geq n_0$, $\lvert \nu_c^{(n)}(t) \rvert \leq 1-\varepsilon$,
$\lvert \nu_v^{(n)}(t) \rvert \leq 1-\varepsilon$ and $\lvert \nu^{(n)}(t) 
\rvert \leq 1-\varepsilon$ (see Feller \cite[Chapter XV.1, Lemma 4]{fell71}). 
From this it follows that
\begin{eqnarray*}
\lefteqn{\left|\frac{\phi_n(\vartheta_n+it)}{\phi_n(\vartheta_n)}\right|} \\
& = & \left|\frac{\left(M_c(q_n (\vartheta_n+it))\right)^{n_c} 
\left(M_v(q_n (\vartheta_n+it))\right)^{n_v} M(z_f (\vartheta_n+it))}
{\left(M_c(q_n \vartheta_n)\right)^{n_c} 
\left(M_v(q_n \vartheta_n)\right)^{n_v} M(z_f \vartheta_n)}\right|\\
&=& \left| \left(\nu_c^{(n)}(t)\right)^{n_c} \left(\nu_v^{(n)}(t)\right)^{n_v} \nu^{(n)}(t) \right|\\
&=& o(1/\sqrt{n})
\end{eqnarray*}
as $n\to\infty$ for all $t \neq 0$ (compare the argument leading to \cite[Chapter XVI.6,
Equation (6.6)]{fell71}), which guarantees (\ref{e:condition3}).$\hspace*{\fill}\square$
\end{proof}

In view of Lemma \ref{condokay}, we may approximate the probability of T-cell 
activation as
\begin{equation}\label{G:Akt_w}
\sP \left( G(z_f) \geq g_{\rm act} \right) \approx 
\frac{\re^{-n I_{n}(a)}}{\vartheta_n \sigma_n \sqrt{2\pi n}} \; ,
\end{equation}
where $G(z_f)$ is the original stimulation rate of (\ref{G:ZV_Trigger_fremd}),
$g_{\rm act}=g_{\rm act}(n)=a n$,
and we assume that $n$ is large enough for a good approximation. (Note that the threshold
$g_{\rm act}$ is not known; but we do know that reactions of T-cells are rare events. Thus, we may
assume that $g_{\rm act}$ is such that large deviations results are applicable, as will also be confirmed
by our simulations below.) The expression in the right-hand side must be evaluated numerically (we used 
Mathematica$^{\text{\textregistered}}$ \cite{wolf03}), since already the moment 
generating functions in (\ref{e:m1}) and (\ref{e:m3}) are unavailable analytically, 
and this carries over to $\vartheta_n$, $\sigma_n^2$, and $I_{n}(a)$  in 
(\ref{e:theta})-(\ref{e:I}). 

\begin{figure}[h]
\begin{center}
\subfigure[$z_f=0$, linear scale]{\includegraphics[width=.45\textwidth]{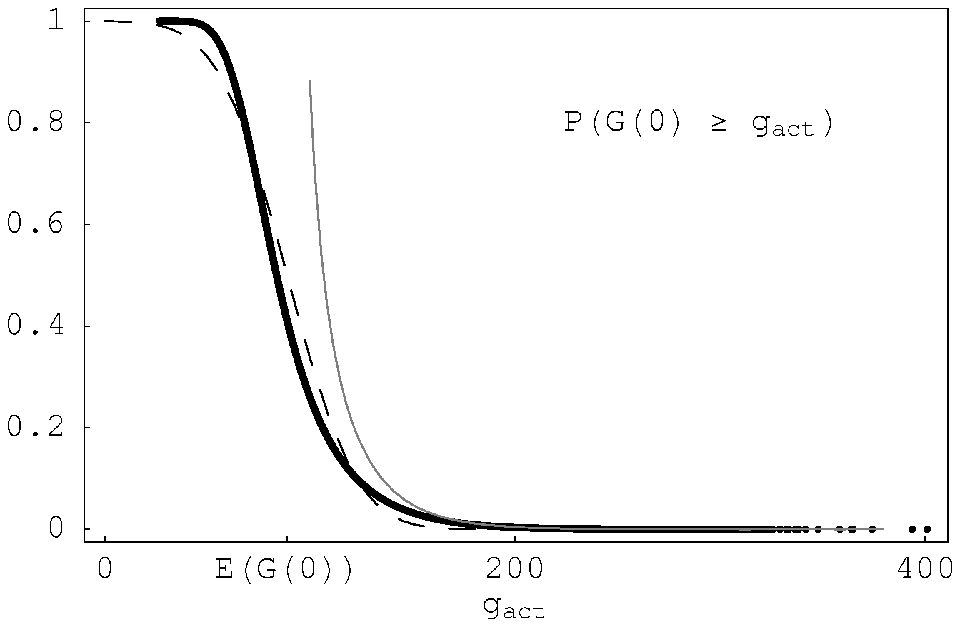}} 
\quad
\subfigure[$z_f=0$, logarithmic scale]{\includegraphics[width=.45\textwidth]{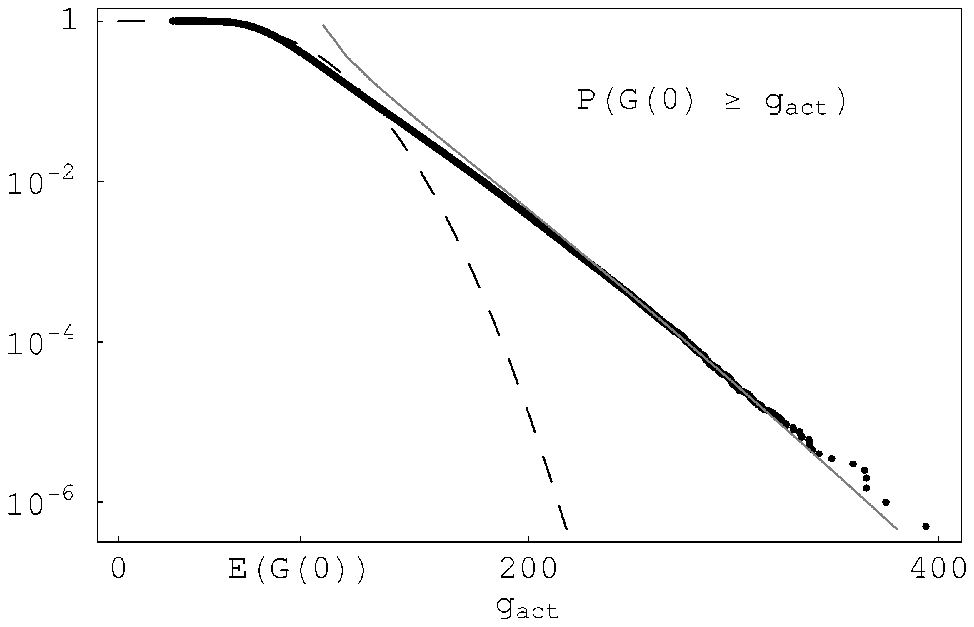}} \\
\bigskip
\subfigure[$z_f=2500$, logarithmic scale]{\includegraphics[width=.45\textwidth]{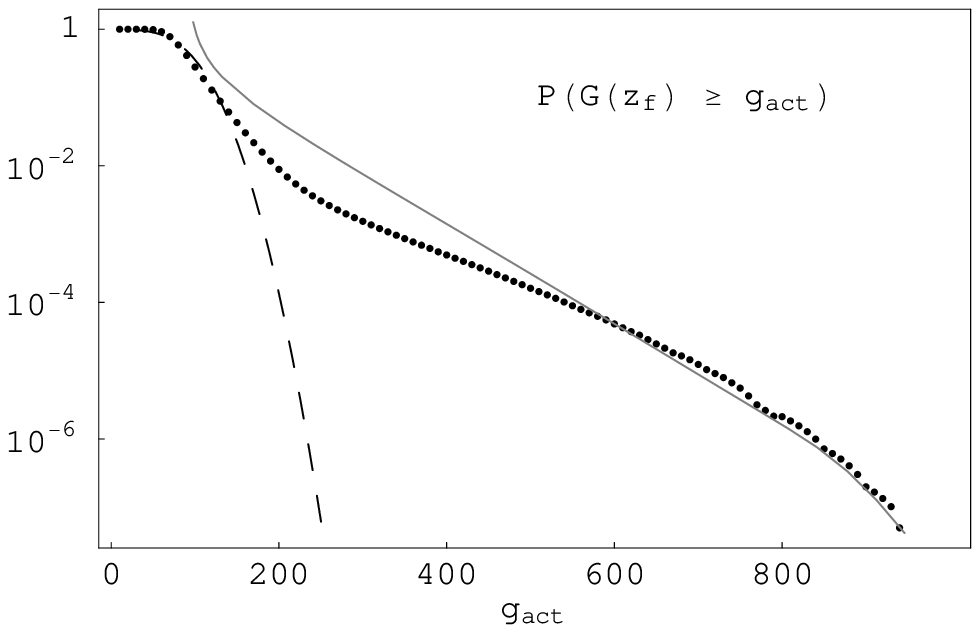}}
\caption{$\sP(G(z_f) \geq g_{\rm act})$ as a function of $g_{\rm act}$, 
for the self background
($z_f=0$), and a very large number of foreign peptides
($z_f=2500$). The thick black curve and the black points, respectively,
form the simulated distribution of two million [(a),(b)]
and twenty million [(c)] sampling points. The dashed curve is the normal approximation,
and the grey curve is the large deviation approximation (\ref{G:Akt_w}).
The simulation in (c) was kindly provided by F.\ Lipsmeier.}
\label{f:comparison}
\end{center}
\end{figure}

Let us consider the activation curve for two extreme cases, namely,
the  self-background ($z_f=0$), and a very large number of foreign peptides
($z_f=2500$). Fig.\ \ref{f:comparison} shows the simulated curve 
in comparison to the normal approximation and the approximation in (\ref{G:Akt_w}).
As was to be expected, the normal approximation describes the central part well, whereas 
for the right tail (the relevant part of the distribution for the problem at hand) the 
large deviation approximation is appropriate. For $z_f=0$, the latter describes the simulated 
distribution in an excellent way; for $z_f=2500$, it still gives
correct approximations beyond $g_{\rm act}=550$, which is the region we are interested in (see the next subsection).
An improved approximation of the entire curve 
is obtained in BRB \cite{vdb01} by combining the normal and
the large deviation approximations, applying them to the self-peptides only,
and performing a convolution with the single foreign one.  We prefer the 
direct approach (\ref{G:Akt_w}) here, because it makes the large deviation aspect more transparent, 
and because it generalizes easily to situations with more than one foreign peptide.
In fact, rather than taking the limit in the way described above, we could as well 
consider a sequence of models with $n_f$ different foreign peptides and let $n \to 
\infty$ such that $n_c/n$, $n_v/n$ and $n_f/n$ each tend to a constant; the approximation 
of our given finite system by (\ref{e:theta})-(\ref{G:Akt_w}) would remain unchanged.

\subsection{Activation curves without negative selection}
\label{s:actneg}

\begin{figure}[h]
\begin{center}
\subfigure[exponential distribution]{\includegraphics[width=.45\textwidth]{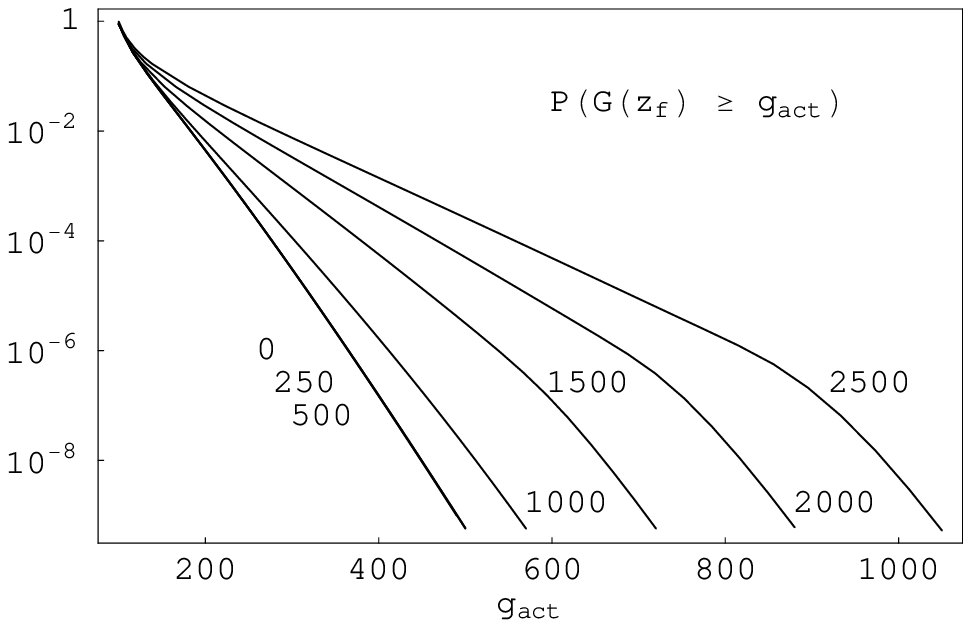}} 
\quad
\subfigure[log-normal distribution]{\includegraphics[width=.45\textwidth]{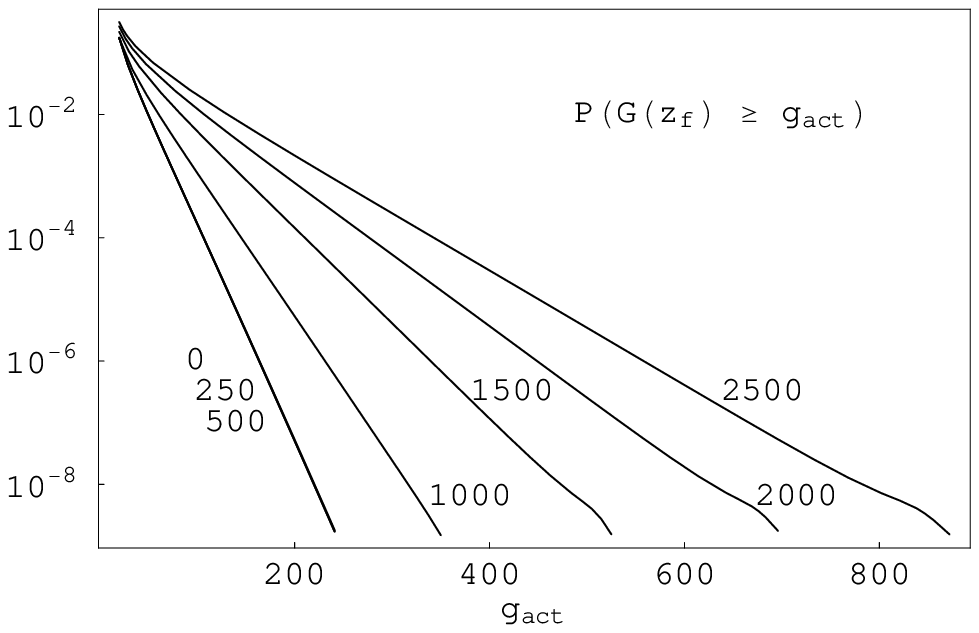}}
\caption{\label{f:akt_without_sel} Activation curves for values of $z_f$ ranging from
$0$ to $2500$, calculated according to approximation (\ref{G:Akt_w}) for two different distributions
for $\mathcal{T}_{ij}$.
The horizontal axis is chosen to start at a value of $g_{\rm act}$
that yields a probability close to 1 in this approximation.}
\end{center}
\end{figure}

As we have seen in the last subsection, the approximation in (\ref{G:Akt_w})
is suitable for the calculation of the activation curves for various values of 
$z_f$. Fig.\ \ref{f:akt_without_sel} shows the curves as a function of $g_{\rm act}$
for exponentially (a) and log-normally (b) distributed $\mathcal{T}_{ij}$.
The results in (a) and (b) are qualitatively the same. Namely,
we observe that the curves for $z_f=250$ and $z_f=500$
(both $\leq \sE(Z_1^{(c)})=500$) do not differ visibly from the curve for the 
self-background; but, for $z_f>1000$ and $g_{\rm act}>500$, condition (C1) 
is fulfilled. Therefore the model can indeed explain how T-cells are able to 
distinguish between self and non-self. Comparison with Fig.\ $3$ of BRB \cite{vdb01} 
shows that the separation of the activation curves is indeed similar to that 
in the original model. In terms of the cartoon in Fig.\ \ref{f:illustration}, the 
threshold value $g_{\rm act}$ can be chosen so that T-cell 2 will be activated 
when it encounters APC 2 with three foreign peptides (the circles), while the other 
APCs without foreign peptides (the non-circles) will not activate any T-cell. 

The intuitive reason behind the self versus non-self distinction is an elevated 
number of presented foreign peptides in comparison with the copy numbers of individual
types of self-peptides. 
Indeed, this increases the variability of $G$ (which is reminiscent of the fact that 
for $n \geq 2$ i.i.d.\ random variables with positive variance, $nY_1$ has a larger variance than 
$\sum_{i=1}^n Y_i$). So far the number of presented foreign peptides has to be 
fairly large: at least as large as the copy number of constitutive ones, which are, 
in turn, more abundant than the variable ones (one
might actually reformulate the hypotheses pair
\eqref{eq:H_0} and \eqref{eq:H_A} so as to
test whether the 
foreign antigen is  more abundant than the constitutive peptides
or not).
However, this restriction vanishes 
when we take the training phase of the young T-cells into account, as will be done 
in the next subsection.

\subsection{Activation curves with negative selection}

In order to model the process called {\em negative selection}
(see Jiang and Chess \cite{jiang06} for a recent review of the
biological details), one postulates 
a second 
threshold $g_{\rm thy}$ with a similar role as $g_{\rm act}$. If the stimulation 
rate of a young T-cell in its maturation phase in the thymus (where the APCs only 
present self-peptides) exceeds this threshold, then the T-cell is induced to die. 
For a caricature version of this process, let us assume that T-cell types present
in the thymus exist in one copy each and encounter exactly one APC there.

The model then consists of two parts: first, the maturation phase is modelled to 
characterize the T-cell repertoire surviving negative selection; second, activation 
curves are calculated for this surviving repertoire. In the first step, we have 
to calculate the probability to survive negative selection conditional on the 
type of the T-cell:
\[
\sP(\text{survival of a T-cell of type } i)
=\sP\left(\sum_{j \in \mathcal{C}_i} Z_j^{(c)} w(\tau_{ij}) +
\sum_{j \in \mathcal{V}_i} Z_j^{(v)} w(\tau_{ij})<g_{\rm thy} \right)\; .
\]
In this case the conceptual difference between $\mathcal{C}_i$ and $\mathcal{V}_i$
has an effect, which is essential. The constitutive types of peptides are the same on 
each APC, both in the thymus and in the rest of the body. Only these can be ``learnt'' 
as self by negative selection -- the $\mathcal{V}_i$, being a fresh sample for every 
APC, are entirely unpredictable. Therefore we have 
\[
\sP(\text{survival of a T-cell of type } i)=\sP({\rm survival}\, 
| \, W_{ij}=w(\tau_{ij}) \, \forall j \in \mathcal{C}_i)\; . 
\]
But in the case of fixed constitutive copy numbers $z_c$, the constitutive part of the 
stimulation rate reads $G^{(c)}=\sum_{j \in \mathcal{C}_i} z_c w(\tau_{ij})$, which 
is {\em constant} for fixed $i$. Therefore we have 
\[
\sP({\rm survival}\, | \, W_{ij}=w(\tau_{ij}) \, 
\forall j \in \mathcal{C}_i)=\sP({\rm survival}\, | \, G^{(c)}=g_i)\; .
\]
This simplifies the second step, the calculation of the activation curves conditional 
on survival: only a single integration step is required. Numerically, it turns out
that (C1) is already fulfilled for $z_f \leq 500$ ($=\sE(Z_1^{(c)})$).
Actually, the detection threshold for foreign antigens is reduced drastically (to about a third of 
the original value).

In our case (where the copy numbers vary from APC to APC),
$G^{(c)}=\sum_{j \in \mathcal{C}_i} Z_j^{(c)} w(\tau_{ij})$, the constitutive part,
varies from encounter 
to encounter. Indeed, whereas the $w(\tau_{ij})$ are fixed for each T-cell, the
copy numbers are tied to the APCs. Therefore $\sum_{j \in \mathcal{C}_i} w(\tau_{ij})$
is not sufficient to determine $G^{(c)}$, and hence the survival probability; rather, 
the entire collection of the individual stimulation rates $w(\tau_{ij})$ for the 
constitutive types must be known to calculate the probability of the young T-cell 
to survive negative selection. The corresponding convolution required in the second step
involves high-dimensional  integrals, which appear to be computationally infeasible.
In Van den Berg and Molina-Paris \cite{vdbm03} this difficulty is tackled by simplifying
the distribution of $W$ to a Bernoulli variable. Here we resort to simulations.

\begin{figure}[h]
\centerline{\epsfig{file=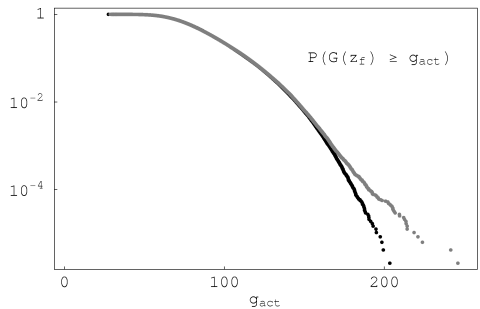,width=65mm}}
\caption{\label{f:akt_with_sel} Simulated activation curves 
with $g_{\rm thy}=140$ for $z_f=0$ (black curve) and $z_f=500$ (grey curve)}
\end{figure}

To this end, we assume that each mature T-cell encounters the same number
(in our simulation $1$) of APCs in the rest of the body. For $g_{\rm thy}=140$ 
(which for our choice of parameters corresponds to thymic deletion of about $5\%$ 
of the young T-cells) the activation curves are shown in Fig.\ \ref{f:akt_with_sel}. 
As in the case of fixed copy numbers, we observe an incipient separation of the 
activation curves for $z_f=0$ and $z_f=500$. All in all, the above shows that the 
reduced detection threshold for foreign antigens occurs in the case of random copy 
numbers too. However, it seems that the separation of the activation curves is 
less pronounced here than in the case of constant copy numbers. This is plausible 
because the copy numbers of several constitutive peptides could be large (comparable 
to the copy number of the foreign peptide) and therefore the recognition does not 
work equally well. More pronounced separation of the curves 
could be achieved by
including mechanisms of so-called peripheral tolerance 
into the model. These mechanisms control the immune response
of T-cells once they have left the thymus (mainly by the action of
T-suppressor cells); see \cite{jiang06} for review and \cite{vdbm03}
for modelling approaches.

\section{Discussion} \label{s:discussion} 

We have analysed mathematically how T-cells can use large deviations to 
recognize foreign antigens. Before we discuss details of our results, 
let us pause here to discuss the notion of recognition that emerges from 
this analysis.

As explained in the introduction, the task the immune system faces is
to recognize foreign antigens against a background of self-molecules.
In much of the biological literature (and in line with intuitive understanding), 
this recognition is implied to be specific on a molecular basis. In sharp 
contrast, the mechanism proposed by BRB \cite{vdb01}, and further analysed 
here, dispenses with such a high degree of strict specificity and, instead, 
is based on elevated copy numbers of the foreign antigen, relative to the 
self-background. This alleviates the Mason paradox about the repertoire size 
versus the universe of potentially relevant peptides. In this sense, specific 
recognition of antigen types is replaced by 
{\em probabilistic recognition}
(this concept also appears elsewhere in biology, for example in the
olfactory system; see, e.g., Lancet et al.\ \cite{lancet93}).
We have examined this fact in detail for the situation without negative 
selection, where the principle becomes most transparent and can be largely 
dealt with via a large deviation analysis. Negative selection basically lowers 
the relevant background stimulation rate. Therefore it is easier for the 
foreign peptide to stand out against it, but the basic principle of recognition 
on the basis of frequencies remains the same.

The generalization we have examined in this article concerns precisely this 
fundamental issue of frequencies. In the original BRB-model \cite{vdb01}, these
frequencies were considered fixed (within the variable and constitutive class). 
In a subsequent paper, \cite{vdbm03}, a more sophisticated model of copy number
variability was introduced (to reflect the fact that different cell types produce
different proteins in different amounts, and some specialized cells may even 
produce large amounts of a variable peptide). We have replaced this here
by a simpler approach that makes the analysis more amenable.

At the same time, allowing the $Z_j$ to vary allows us to weaken assumption (A1), 
which is somewhat too restrictive in that it assumes that, for all $j \in 
\mathcal{P}_i$, all pMHCs are in the perfectly bound state. But if we simply 
reinterpret $Z_j$ as the (fluctuating) number of {\em bound} antigens rather 
than the total number (i.e., corresponding to $\hat c_{ij}$ rather than to $z_j$), 
then we can circumvent (A1) and, at the same time, free ourselves from the 
details and limitations of the deterministic binding kinetics, the assumptions 
of which may sometimes be violated, for example, if copy numbers are low.

Although, in this vein, a detailed model of the binding kinetics may
not be required, one last remark is in order concerning a potential
{\em stochastic} model. Stochastic models of binding kinetics, e.g.\ as 
described in Ethier and Kurtz \cite[Chapter $11$]{ethi86}, take care of 
the finite number of molecules, and of the resulting fluctuations of bound 
molecules over the short time scales of the binding kinetics. Over the 
longer time scales to be considered for our averaged stimulation rates, the 
corresponding time averages would be relevant. In particular, these would in 
general not be integers; in this light, the factor $q$, which turns the
multipliers of the sums in (\ref{G:ZV_Trigger_fremd}) into non-integer values, 
is rendered plausible.

Summarizing, our results show that the {\em probabilistic} recognition phenomenon is 
astonishingly robust against fluctuations of the copy numbers. Similarly, 
generalizations in various other directions that take into account many more 
biological details (see e.g.\ \cite{vdbm03,vdbr03,vdb04}), have shown that 
the phenomenon persists in a robust way. One may therefore hope that several 
details of the assumptions may actually be dispensed with. Future work will 
thus aim at identifying the joint mathematical content of the underlying family 
of models. In particular, this will include an analysis of what aspects of the 
distribution of the $W$'s are key to the observed phenomenon. Luckily, the 
scope of the large deviation result Theorem \ref{T:verall_B_R}
is wide enough to cope with very general situations. In particular, one is not 
bound to sums of independent random variables, but can also tackle 
dependencies (like, for example, those introduced by negative selection) 
within the present framework. Furthermore, the large deviation framework may
also be used to construct efficient simulation methods for the tail events,
as will be described in a forthcoming paper (F.\ Lipsmeier \& E.\ Baake,
in preparation).


\begin{acknowledgements}
It is a pleasure to thank Hugo van den Berg for explaining the details of the 
BRB-model and providing helpful comments on the manuscript, and Michael Baake 
for extensive discussions. The simulation results in Fig.~\ref{f:comparison}(c)
were kindly provided by Florian Lipsmeier. This work was supported by the Bilateral Research
Group ``Random Spatial Models in Physics and Biology'', funded by DFG and NWO.
\end{acknowledgements}


\end{document}